\documentclass[useAMS,usenatbib]{mn2e}
\usepackage[dvips]{graphicx}
\title{Linking halo mass to galaxy luminosity}
\author[A. Vale and J. P. Ostriker]{A. Vale$^{1}$\thanks{E-mail: avale@ast.cam.ac.uk} and J. P. Ostriker$^{1}$\\
$^{1}$Institute of Astronomy, University of Cambridge, Madingley Road, Cambridge CB3 0HA, United Kingdom}
\date{Accepted...... Received......; in original form 20 February 2004}
\begin{document}
\maketitle

\begin{abstract}
In this paper we present a new, essentially empirical, model for the
relation between the mass of a dark matter halo/subhalo and the
luminosity of a galaxy hosted in it.  To estimate this, we replace the
assumption of linearity between light and mass fluctuations with the
assumption of monotonicity between galaxy light and halo or subhalo
mass. We are enabled to proceed with this less restrictive ansatz by
the availability of new, very high resolution dark matter simulations
and more detailed and comprehensive global galactic luminosity
functions.

We find that the relation between halo/subhalo mass and hosted  galaxy
luminosity, is fairly well fit by a double power law. That between
halo mass and group luminosity has a shallower slope for an
intermediate mass region, and is fairly well fit by a two branch
function, with both branches double power laws. Both relations
asymptote to $L\propto M^4$ at low $M$,  while at high mass the former
follows $L\propto M^{0.28}$ and the latter  $L\propto M^{0.9}$.

In addition to the mass-luminosity relation, we also derive results
for the occupation number, luminosity function of cluster galaxies,
group luminosity function and multiplicity function. Then, using a
prescription for the mass function of haloes in under/overdense
regions and some further assumptions on the form of the mass density
distribution function, we further derive results for biasing between
mass and light and mass and galaxy number, light distribution function
and the void probability distribution.

Our results for the most part seem to match well with observations and
previous expectations. We feel this is a potentially powerful way of
modelling the relation between halo mass and galaxy luminosity, since
the main inputs are readily testable against dark matter simulation
results and galaxy surveys, and the outputs are free from the
uncertainties of physically modelling galaxy formation.
\end{abstract}


\section{Introduction}

In recent years, N-body numerical simulations have given us a good
understanding of dark matter structure for standard cosmological
scenarios, while large scale observational surveys have done the same
for the distribution of galaxies. In this way, we are now developing a
good picture of how mass and luminosity are distributed in the
universe. However, it is still not well known how to connect the two
pictures. While it is well established that dark matter haloes are the
hosts of the observed galaxies, it is still poorly understood how the
former are related to the latter. Further, the picture is complicated
by the fact that what is usually taken as a halo in simulations would
often host multiple galaxies, especially for higher masses. To analyse
the issue fully, it is necessary to look at the halo substructure,
since each subhalo can host a galaxy.  Establishing such a link
between halo mass and galaxy luminosity would be important because,
first of all, it would allow us to have a direct connection between
theory and observation, dark matter haloes and galaxies. Further, it
could also shed some light into the theory of galaxy formation.

There are several ways in which this problem can be studied. The more
direct ones involve either numerical simulations including gas
dynamics \citep{whs,ytj,pjf,nfco,Berlindetal}, or semi-analytical
models of galaxy formation
\citep{kns,gbf,kcda,kcdb,bba,bbb,sd,sls,wsb,bfb,Berlindetal}, but,
while they explicitly give the properties of galaxies located in a
given halo, they have the added difficulty that many of the mechanisms
involved in galaxy formation are poorly understood, and difficult to
compute. Their complexity could also mask any fundamental relations
that might be present between halo and galaxy properties.

More indirect approaches have also been studied. The halo occupation
distribution (HOD) model
\citep{seljak,benson,bws,zt,BerlindWeinberg,Berlindetal,mp} is based
on the probability $P(N|M)$ that a halo of mass M is host to N
galaxies.  By specifying the $P(N|M)$ function, along with some form
for the distribution of dark matter and galaxies within each halo, it
is then possible to relate different statistical indicators of the
dark matter and galaxy distributions, such as correlation functions,
to each other. This fully specifies the bias between the galaxy and
the underlying matter distributions.  A recent paper by
\citet{Kravtsovetal}, has done a detailed study of results from
simulations and related them to the HOD model, and has concluded that
the form of $P_s(N_s|\mu)$ for the subhaloes is approximately
universal, where $\mu$ is the subhalo mass scaled to an appropriate
minimum mass.  This paper also give results for the relation between
galaxy absolute magnitude and halo circular velocity.  Other work
\citep{BoschYangMo,YangMoBosch,Moetal} has taken this approach one
step further by studying not only the number of galaxies associated
with each halo, but also their luminosity, by building the conditional
luminosity function, $\Phi(L|M){\rm d}L$. This gives the number of
galaxies with luminosities in the range $L\pm{\rm d}L/2$ contained in
a halo of mass $M$. While this work directly relates the halo mass to
the galaxy luminosity, it does so only to the average values, lacking
the full statistical treatment which is analysed in the HOD models.

Other authors have used a slightly different method. Instead of trying
to specify the number of galaxies in each halo, they treat the halo as
a whole and identify it with a galaxy group. Then, by comparing the
group luminosity function with the halo mass function, they obtain the
luminosity associated with each halo
\citep{PeacockSmith,MarinoniHudson}, and also develop ways to estimate
the number of galaxies hosted in a halo, thus coming back partly to
the $P(N|M)$ estimate of the HOD models.

In the present paper, we follow a new and conceptually clear approach
based on one simplified and testable hypothesis: there is a one to
one, monotonic correspondence between halo/subhalo mass and resident
galaxy luminosity. We might call this the empirical (rather than the
semi-analytical) approach because there is no attempt to physically
model the galaxy formation process. Instead we take from observations
the galaxy luminosity distribution and match it with the theoretical
halo/subhalo distribution. This has the additional advantage of
naturally giving a lower mass threshold for haloes that host luminous
galaxies, as the luminosity decreases sharply with mass for less
massive haloes. It also implicitly gives rise to galaxy systems, if
one identifies a system of a massive halo and its subhaloes with the
central galaxy and its satellites in groups and clusters. We find that
the single assumption is very powerful and allows us to compute, and
compare to observations many quantities from bias to the void
distribution function to the spatial correlation function.

This paper is organized as follows: in section 2 we present our model
for the subhalo mass distribution in a parent halo, and build the
global subhalo mass function. In section 3, we derive the relation
between mass and luminosity, as well as some other functions such as
the luminosity function of cluster galaxies, the group luminosity
function and the multiplicity function. In section 4 we study how to
apply the relation we obtain to get the light density and the number
density of galaxies as a function of mass density, and also obtain
results for the distribution function of light density and the void
probability function. Finally, we conclude in section 5.

Throughout we have used a concordance cosmological model, with
$\Omega_m=0.3$, $\Omega_\Lambda=0.7$, $h=0.7$ and $\sigma_8=0.9$
\citep{triangle,wmap}.

\section{Subhalo Distribution}

The first step in building our model is to specify the mass
distribution of subhaloes for a given parent halo. We adopt the
following function:

\begin{equation} \label{nsubhalo}
N(m|M) dm = A \Big(\frac{m}{x \beta M}\Big)^{-\alpha} {\rm
exp}\Big(-\frac{m}{x \beta M}\Big) \frac{dm}{x\beta M} \, ,
\end{equation}

\noindent which gives the number of subhaloes with masses in the range
$m$ to $m+dm$, for a parent halo of mass $M$. The normalisation $A$ is
such that the total mass in these subhaloes, $\int_{0}^{\infty}m
N(m|M) dm$, is a fraction of the parent halo mass, $x \gamma M$ (where
the factor $x$ accounts for the added mass of the original,
unstripped, subhaloes).  With this definition, we can write $A$ as

\begin{equation} \label{normalization}
A=\frac{\gamma}{\beta\Gamma(2-\alpha)} \, .
\end{equation}

This expression is motivated by recent analysis of high resolution
dark matter simulations of \citet{jochen},
and we use for the parameters the values $\alpha=1.91$, $\beta=0.39$
and $\gamma=0.18$.  Its results are similar to those obtained by the
simulations of \citet{lucia}, who find a power law fit to the subhalo
mass function with a slope close to -2 (that is, in terms of equation
\ref{nsubhalo}, $\alpha=2$). However, they also find that if they only
include the lowest mass bins where statistical errors are smallest,
this slope is reduced to values around -1.9, very similar to what we
use here.   It further agrees well with the cumulative mass function
derived analytically by \citet{ol} (see section below on the
occupation numbers for further discussion), and it is also similar to
the power law form for the subhalo number given in \citet{ShethJain},
only instead of having a sharp cutoff at the parent halo mass, we
introduce a smooth exponential cutoff from a mass $\beta M$.  Since it
is a Schechter function, it is also similar to the halo mass function,
and the slope  $\alpha$ is close to the expected value of the slope of
the halo mass function.

It gives a total mass in subhaloes of 18\%, close to but slightly
higher than the values of $\approx10\%$ obtained in different studies
(see, for example, \citealt{TormenDiaferioSyer,Ghignaetal}).
\citet{lucia} derive a lower value of around 6\% at radius $r_{200}$
(where the average overdensity is 200 times the critical), though they
also find that this fraction can be as high as 10-15\% in some
cases. However, $r_{200}$ is smaller than the virial radii typically
measured from the simulation results by \citet{jochen}, which helps to
explain the difference.

It is worth noting that, like previous results from simulations and
analytical modelling (e.g., \citealt{moore, Kravtsovetal,lucia,ol}),
the shape of the subhalo mass function given by equation
\ref{nsubhalo} is independent of parent halo mass.  This does not
mean, however, that one should automatically expect the same to be
true of satellite galaxy distribution in galaxies like the Milky Way
and clusters. In fact, as can be see from our results further below,
the mass luminosity relation has very different behaviour depending on
the mass of the host halo/subhalo; this leads to parent halos of very
different masses (such as a cluster and a galaxy) having very
different satellite luminosity distributions, even though the scaled
mass functions of their subhaloes are identical.

A very important point when analysing this expression is to note
exactly what mass is being accounted by $m$. In fact, the distribution
is valid for the present mass of the subhalo satellites, obtained when
$x=1$, which is measured after the tidal stripping of the outer parts
of the subhalo in the halo potential well. However, to use this
function to build the host distribution and to subsequently compare it
to the galaxy luminosity function, it is necessary to use the
original, unstripped mass of the subhalo, since only then can we
establish a monotonic correspondence between host halo mass and galaxy
luminosity. The factor $x$ put into the expression takes this mass
loss into account, since we can then treat the mass $m$ in equation
(\ref{nsubhalo}) as the original mass, with the stripped mass being
then $m/x$ with $x>1$. Obviously, the actual factor for a given
subhalo will be highly variable, but for reasons of simplicity and
also because it is not well known (see e.g. \citealt{Hayashietal} for
a treatment of the profile of stripped subhaloes), we will be treating
this as an average applicable to all subhaloes, and using $x=3$ in the
present work. It should be noted that, since we are taking the total
subhalo mass to be 18\% of the parent halo mass, this factor cannot be
more than 5, assuming that the majority of the mass in the parent halo
was built up by stripping of the subhaloes. A recent paper by
\citet{kravtsovdwarfs} includes a comparison between maximum and
present mass of subhaloes in the dwarf mass range. Their results (see
their figure 4) seem to support that the average change in mass does
not depend on the mass of the subhalo, and an average mass stripping
factor of 3 as we adopt here seems a fair agreement.

All plots and further expressions (where we will drop the factor $x$)
presented refer to this original, unstripped mass; to invert our
approximation is simply a case of dividing the subhalo mass by a
factor of 3 to represent the actual stripped mass which would be
measured in a simulation. Ideally, we should use instead a
distribution for the maximum circular velocity of the subhaloes.  Even
though this quantity also shows a large scatter between the maximum
value and that at present after stripping \citep{kravtsovdwarfs}, its
relative change should be smaller than what would be expected for a
given change in mass. This should be the case especially for subhaloes
massive enough to host galaxies, which would have only their outer
parts stripped off, and where therefore the inner parts which
determine the peak circular velocity are left relatively undisturbed.
This would mean that a calculation based on the cumulative number
function like we use here would incurr a smaller error, even with a
large scatter, and in later work we will transform to  that variable.

Using the expression for the subhalo number (\ref{nsubhalo}) together
with the halo mass function, $n_h(M)$, it is then possible to build
the global subhalo mass function, that is, the number density of
subhaloes in a given mass range, by

\begin{equation} \label{intsubhalomf}
n_{sh}(m) = \int_{0}^{\infty} N(m|M) n_h(M) dM \, .
\end{equation}

If we assume that the halo mass function has a Schechter form,

\begin{equation} \label{schechterm}
n_h(M) dM = C \Big(\frac{M}{M_*}\Big)^{-a} {\rm
exp}\Big(-\frac{M}{M_*}\Big) \frac{dM}{M_*} \, ,
\end{equation}

\noindent it is possible to write an analytical expression for it:

\begin{equation} \label{subhalomf}
n_{sh}(m) = \frac{2 C \gamma}{M_* \beta^2 \Gamma[2 - \alpha]}
\Big(\frac{m}{\beta M_*}\Big)^{-\frac{a+\alpha}{2}} K_0\Big(2 \sqrt{\frac{m}{\beta M_*}}\Big) \, ,
\end{equation}

\noindent where $K$ stands for the modified Bessel function.

\begin{figure}
\includegraphics[width=84mm]{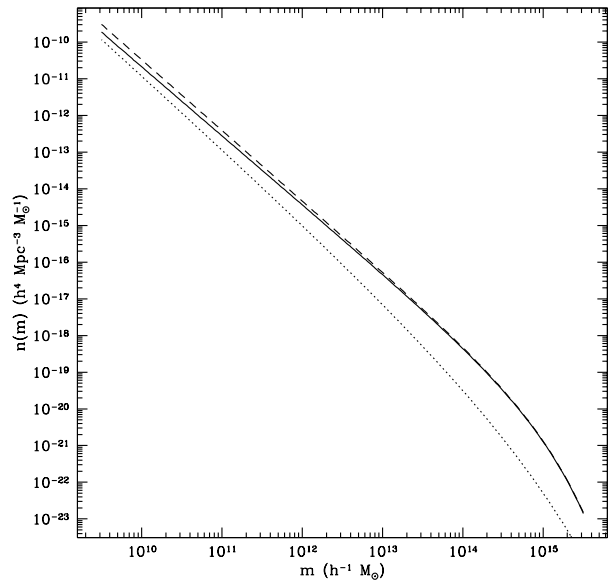}
\caption{Mass function for haloes (solid line) and subhaloes (dotted
line). The dashed line is the sum of the two. The parent halo mass
function is the one proposed by Sheth Tormen (equation \ref{stmf}),
while the subhalo one is the one obtained using our model for the
subhalo number distribution with parent halo mass.} \label{massffig}
\end{figure}

With more general mass functions, the calculation has to be done
numerically. In the present work, we use the more accurate Sheth
Tormen mass function \citep{ShethTormen99},

\begin{equation} \label{stmf}
n_h(M) dM = A \Big( 1+\frac{1}{\nu^{2q}}\Big) \sqrt{\frac{2}{\pi}} \frac{\rho_m}{M} \frac{d\nu}{dM} {\rm exp}\Big(-\frac{\nu^2}{2}\Big) dM\, ,
\end{equation}

\noindent with $\nu=\sqrt{a}\frac{\delta_c}{D(z) \sigma(M)}$,
$a=0.707$, $A\approx 0.322$ and $q=0.3$; as usual, $\sigma(M)$ is the
variance on the mass scale $M$, $D(z)$ is the growth factor, and
$\delta_c$ is the linear threshold for spherical collapse, which in
the case of a flat universe is $\delta_c=1.686$. The result for the
global subhalo mass function is shown in figure \ref{massffig} , where
the Sheth Tormen mass function is also shown.

\subsection{Occupation number}

With the expression for the subhalo number, equation (\ref{nsubhalo}),
it is also possible to calculate the halo occupation number, that is,
the number of subhaloes in a halo of mass $M$:

\begin{equation} \label{occupation}
N_s=\frac{\gamma}{\beta\Gamma(2-\alpha)}\Gamma(1-\alpha,\frac{m_{min}}
{\beta M}) \, ,
\end{equation}

\begin{figure}
\includegraphics[width=84mm]{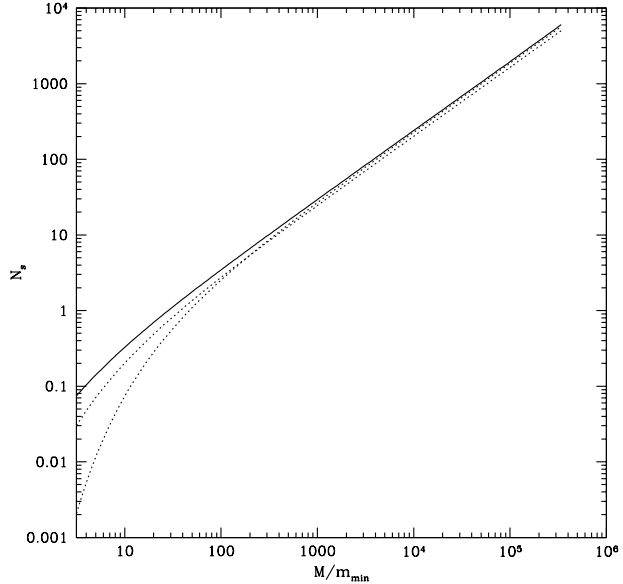}
\caption{Number of subhaloes in a halo of mass M (equation
\ref{occupation}, solid line).The halo mass is scaled to the minimum mass
considered for the subhaloes. The solid line shows the results for
the model used in this paper ($\beta=0.39$). The two dotted lines
show the result of changing the subhalo cutoff (respectively for
$\beta=0.2$ and $\beta=0.05$).} \label{haloocupfig}
\end{figure}

Because the integral diverges, it is necessary to set a minimum
threshold for the subhaloes. This can be set as the minimum mass for a
halo to host a galaxy, in which case this occupation number
corresponds to the number of galaxies contained in a given halo.  This
is one of the basic ingredients of HOD models, where it is taken as
the average number of galaxies in a halo. Figure \ref{haloocupfig}
shows the occupation number as a function of the parent halo mass, in
units of the minimum mass considered. The most prominent feature is
the cutoff for $M/M_{min}\la 5$; this is inherent to our model, since
we consider the total mass in subhaloes to be only 18\% of the halo
mass, and we also have a cutoff in the number of subhaloes, given by
the parameter $\beta$.  At the high mass end, the function is well fit
by a power law, with $N_s\propto M^{0.91}$ for $(M/M_{min})\ga 1000$,
which is similar to what is predicted by most HOD models (see e.g.,
\citealt{BoschYangMo,mp,Kravtsovetal}).

This behaviour is also analogous to that found in the analytically
derived halo mass function of \citet{ol}, who also find a slope of
roughly 0.9 at high $(M/M_{min})$, but a steeper value close to 1 in
the lower range, similar to our results. The values they obtain are
also similar to ours, with one caveat: they correspond to those we
would get if we took the values for the stripped mass of the
subhaloes, instead of what is plotted in figure \ref{haloocupfig};
this would correspond roughly to diving the numbers we get by a factor
of 3.  Our result is also similar to what has been observed for the
case of galaxies in a cluster, with \citet{Kochaneketal} obtaining a
relation for the number of galaxies with luminosity greater than
$L_*$, $N_g(L>L_*)\propto M_h^{1.1}$, where all quantities are
normalised to the usual mass overdensity of 200.  If we assume that
their $L_*$ galaxy would have a mass of around $10^{12} h^{-1} {\rm
M_\odot}$ (see below for the results we obtain), we also obtain
approximately the same number of galaxies in a $10^{15} h^{-1} {\rm
M_\odot}$ parent halo. The overall shape of our occupation number is
also similar to the results obtained by \citet{Kravtsovetal} in their
simulations, although in their case the cutoff is less pronounced, and
in fact is only noticeable for $M/M_{min}\approx 1$. As referred,
however, our cutoff around $M/M_{min}\approx 5$ is in fact a feature
of our model, and something similar should be present as long as a
cutoff in subhalo mass is introduced and there is an upper limit to
the total mass in subhaloes. The normalisation of the curves is also
similar, with a value of $N_s\approx 30$ for $M/M_{min}\approx
1000$. It should however be cautioned that the way in which the
subhalo mass is accounted may not be exactly the same in both cases
(see discussion above).

\section{Mass Luminosity Relation}

Once we have the total mass function for haloes and their subhaloes,
it is then possible to compare it to the galaxy luminosity function to
obtain a mass luminosity relation. As noted, a few prior assumptions
go into this: first, that the galaxies are hosted individually in the
haloes or subhaloes, and that each contains a single one (or, in the
case of the parent haloes, they have one in their centre, plus the
ones in their subhaloes); and second, that the luminosity is a
monotonic function of the halo mass.  The first assumption is
supported by previous studies of the group mass function. For example,
\citet{Martinezetal} obtain a group mass function from the 2dF
catalogue which matches well with theoretical mass functions like the
Sheth Tormen one, equation (\ref{stmf}), indicating that each of their
identified groups corresponds to a halo.  As for the second,
\citet{Neyrincketal} have shown that subhaloes identified in a set of
simulations have a correlation function and power spectrum that
matches the galaxies in the PSCz survey, which shows it is possible to
understand the spatial distribution of galaxies by identifying
galaxies brighter than a given luminosity with haloes larger than a
certain mass.  This is further justified by studies which show that
the only halo property dependent on the large scale environment is the
mass distribution (e.g., \citealt{LemsonKauffmann}), which coupled
with the current understanding of galaxy formation theories should
imply that we have captured most of the environmental dependence of
the galaxy luminosity.

We take the galaxy luminosity function to have the usual Schechter
form:

\begin{equation} \label{schechter}
\phi(L) dL = \phi_* \Big(\frac{L}{L_*}\Big)^{\alpha} {\rm
exp}\Big(-\frac{L}{L_*}\Big) \frac{dL}{L_*} \, .
\end{equation}

The values of the different parameters are taken from the $b_J$ band
2dF galaxy luminosity function \citep{2dF}, with $\Phi^*=1.61\times
10^{-2} {\rm h^3 Mpc^{-3}}$, $M^*_{\rm b_J}-5{\rm log}_{10}h=-19.66 $
and $\alpha=-1.21$.  Although this fit was determined for the
magnitude range $-16.5>M_{b_J}-5 {\rm log} h>-22$, here we will be
extrapolating its result to higher or lower luminosities as necessary.
The mass luminosity relation is then calculated by setting the
luminosity $L$ of a galaxy hosted in a halo of mass $M$ to be such
that the number of galaxies with luminosity greater than $L$ equals
the number of haloes plus subhaloes with mass greater than $M$:

\begin{equation} \label{mlintegrals}
\int_L^{\infty}\phi(L)dL=\int_M^{\infty}[n_h(M)+n_{sh}(M)]dM \, .
\end{equation}

It should be noted that, because the values for the subhalo mass
function are generally lower than those of the halo one, the subhaloes
make only a small contribution to this expression, only being
important at lower mass values. This guarantees that a possible second
generation of subhaloes (subhaloes of the subhaloes considered here)
would not influence this result much, since their mass function would
have values that much lower than the original halo mass function, and
then only at the lowest masses considered.

It is also possible to obtain the group luminosity associated with
each halo. In order to do this, we first assume that a galaxy system
(which would be either an isolated galaxy, a group or a cluster,
depending on the mass of the halo containing it) can be represented by
a halo and its associated subhaloes. The group luminosity for a halo
of mass $M$ is then simply

\begin{equation} \label{grouplum}
L_g(M)=L(M)+\int_0^{\infty}L(m)N(m|M)dm \, .
\end{equation}

\begin{figure}
\includegraphics[width=84mm]{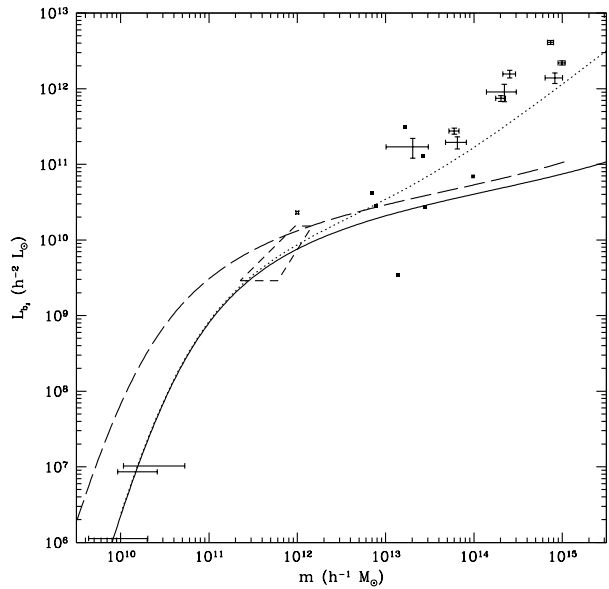}
\caption{Relation between galaxy luminosity and the original mass of the
dark matter halo which hosts it. The solid line is for each
individual halo or subhalo, the dotted line shows the total group
luminosity of the halo plus subhaloes system, in which case the
x-axis $m$ refers to the parent halo mass only. Also shown as a
dashed line is the result when the mass is taken to be the actual
stripped mass of the subhaloes (which is obtained by dividing the
original mass by a factor of 3). The different points are
estimates from different observational studies (see discussion in
text for references).} \label{masslightfig}
\end{figure}

There is an important point that should be considered when analysing
this expression, and that is the possibility of putting a lower limit
in the integral, since there may be a minimum mass for a halo to be
host to a luminous galaxy.  However, as can be seen in figure
(\ref{masslightfig}), which shows the results for the luminosity and
group luminosity associated with a halo, the calculated luminosity
decreases sharply with mass at the low end, becoming negligible for
haloes with masses lower than approximately $10^{9.5} {\rm h^{-1}
M_\odot}$.  Therefore, instead of putting in a cutoff, the value of
which is not clear from what is known of galaxy formation, we simply
use the natural drop off in the derived relation.

This lower limit seems interestingly to be roughly the mass which is
usually considered for the host haloes of dwarf galaxies (for example,
\citet{ThoulWeinberg} give a lower limit of $v_c=30 {\rm km/s}$ to the
maximum circular velocity of halo capable of hosting a dwarf galaxy;
see also \citet{Stoehretal,Hayashietal} for their simulation results
on the halo hosts of dwarf galaxies, where they give similar values
for the less massive of them) , but it is quite low when compared with
the values used in some studies (for example, in the HOD model in
\citet{BerlindWeinberg}), although this may be related to the fact
that most dark matter simulations used in these studies do not reach
such low masses.  Another important aspect that can be seen in the
figure is that the group luminosity only starts to depart from the
halo luminosity for masses above $10^{12} {\rm h^{-1} M_\odot}$. This
is to be expected given the estimated mass of the host haloes of rich
groups and clusters. This difference is the contribution of the
subhaloes, and it is clearly seen that, although their numbers may not
be very significant to the total mass function, they are very
important for the luminosity of high mass haloes, and therefore to the
luminosity function.  Due to their small numbers when compared with
subhaloes, it is to be expected that subsubhaloes make only a small
contribution to this group luminosity, and that this would only be
noticeable for the most massive parent haloes, where the subsubhaloes
can be massive enough to be luminous.  Noticeably, the group
luminosity has a middle region of intermediate mass with a shallower
slope, and then an upturn for high mass, which was also present in
similar studies which used the group luminosity function (e.g.,
\citealt{PeacockSmith}), while the single halo luminosity has a much
shallower high end slope.

We find that the mass luminosity relation we obtain can be fairly well
fit by a double power law of the type:

\begin{equation} \label{fit}
L=A\frac{(m/m')^{b}}{(c+(m/m')^{d k})^{1/k}} \, .
\end{equation}

\noindent For the luminosity of individual galaxies (the solid line in
figure  \ref{masslightfig}), the parameters are $A=5.7\times10^9$,
$m'=10^{11}$, $b=4$, $c=0.57$, $d=3.72$ and $k=0.23$. Therefore, at
the low mass end,  we have $L\propto M^4$. In fact, if we go to very
low luminosities, the slope will actually be slightly steeper;
nonetheless, the region of interest  essentially begins for
luminosities above a few times $10^5 h^{-2}  {\rm L_\odot}$ (which
corresponds to the fainter dwarf galaxies; however, note that such low
luminosities are outside the range of the Schechter fit to the 2dF
luminosity function, so this is based on an extrapolation of the 2dF
results to this luminosity range), and for these the slope is
approximately 4.  Such a slope is actually what you  would expect from
a straightforward comparison between the halo mass function (assuming
a slope of -1.8 at the low mass end) and the galaxy luminosity
function. Since at the low mass end the subhaloes give an important
contribution to the total number of hosts (see figure \ref{massffig}),
this is most likely a coincidence, arising from the fact that the
total host distribution (haloes+subhaloes) has a slope similar to the
halo mass function.  At the high mass end, we essentially obtain the
relation between the halo mass and the luminosity of the brightest
cluster galaxy, which has a much shallower slope ($L\propto
M^{0.28}$). This is most likely due to the fact that, by construction,
the mass term refers to the mass of the entire halo hosting the
cluster, and not just to the mass in the region of the galaxy itself.

The halo mass / group luminosity relation is not really well fit by a
double power law, since it has a third region for middle values of
mass with a shallower slope than either of the asymptotic
values. However, we find that it is possible to describe it as two
different double power law branches, which provide a fair fit to the
results.  Thus, for $m<10^{12} h^{-1} {\rm M_\odot}$, the group
luminosity is essentially the same as the luminosity of the parent
halo (as these haloes do not have subhaloes massive enough to be
luminous), so we fit it also with equation \ref{fit}, and use the same
parameters as before.  For parent halo masses higher than $10^{12}
h^{-1} {\rm M_\odot}$, we find that a simple double power law of the
form $4.8\times10^{10}(m_1^{0.9}+0.6 m_1^{0.4})$, with
$m_1=m/(3.5\times10^{13} h^{-1} {\rm M_\odot})$ is a good fit.  Such a
behaviour, of a curve with a relatively flat slope in the intermediate
mass range and steeper slopes at both the low and high mass ends is
similar to what was observed by \citet{PeacockSmith}, who compared the
AGS group luminosity function with the halo mass function, although
the actual shape of the curve is somewhat different from what we find
here.

At the high mass end, the cluster luminosity is almost directly
proportional to halo mass (in fact, $L\propto M^{0.9}$). This means
that the resulting mass to light ratio will be almost constant, rising
only very slowly with halo mass, which matches well with previous
results for the mass to light ratio of clusters (e.g.,
\citealt{bcdoy,Kochaneketal}).  However, the values we obtain for the
group luminosity seem to be smaller than the observational results.
Further, the value derived by \citet{fhp} for the cluster blue mass to
light ratio is $450 \pm 100 h(M/L)_\odot$, which is roughly consistent
with the value we obtain around $10^{13}-10^{14} h^{-1} {\rm
M_\odot}$, but smaller than what we get at $10^{15} h^{-1} {\rm
M_\odot}$.  Since the mass luminosity relation for a single halo is
dominated at the high end by the parent haloes, it seems unlikely that
this result could be much modified in the scope of our model, since
both the halo mass function and the luminosity function are well
known.  This then means that the problem would lie in the subhalo
distribution for these massive parent haloes; it would be necessary
either to have more of them, or else for them to be more luminous.
The two are in fact related, since an increase in the number of
subhaloes causes an increase in the total number of hosts, and thus a
decrease in the luminosity of a halo (especially in the low mass
range, where subhalo number is more significant).  However, our
results for the occupation number and the luminosity of lower mass
haloes seem to be in good agreement with observations and prior
theoretical models, which leads us to believe that the most likely
cause of this result is the simplistic way in which we treated subhalo
mass stripping.
 
We can also compare our results with those in \citet{BoschYangMo}, who
fit the mass to light ratio in different models they study to a double
power law.  In general, their results are in fair agreement with what
we obtain.  At the low mass end, the results are quite similar, with
these authors obtaining a minimum of the mass to light ratio at a
slightly lower mass.  At the high mass end, they obtain a steeper mass
to light ratio as a function of mass, although this is due to the
flattening off we find in our results, since in the intermediate range
$10^{13}-10^{14} h^{-1} {\rm M_\odot}$ we obtain higher mass to light
ratios and a steeper relation.  This discrepancy is most likely due to
the factor that these authors are fitting the mass to light ratio to a
double power law, which as was discussed above does not provide a good
fit to our results.  In fact, when they adopt a different model with a
fixed mass to light ratio at high mass, their results agree with ours
slightly better in this intermediate region.

Overall, our results seem a fairly good match to estimates of mass
taken from a range of observational results across the entire mass
range. Shown are points for the three most luminous of the dwarf
spheroidals in the local group, where the luminosity was taken from
the review by \citet{Mateo}, and the mass was estimated from the
results of \citet{Hayashietal} (see the results in their figure 13 for
bounds to the peak circular velocity function of unstripped NFW haloes
estimated to be possible hosts of the local group dwarf spheroidals) ,
by assuming that the relation between mass and luminosity is
monotonic.  The dashed box represents the relation obtained from the
weak lensing study of \citet{lensing}, for galaxies around $L^*$
(where we have taken the values applicable to a NFW halo). The results
for poor groups are taken from \citealt{poorgroups}, while those for
clusters are from \citealt{clusters}, where the bounds come from the
two different methods the authors use for estimating fore/background
corrections. Also included is a point for the Milky Way \citep{Allen}.

\subsection{Luminosity function of cluster galaxies}

The major difficulty in obtaining a luminosity function for galaxies
in clusters lies in distinguishing what haloes should be treated as
rich clusters in our model, and which are simply groups. We adopt the
Abell definition for rich cluster, namely that it must have upwards of
30 objects brighter than $m_3+2^m$, where $m_3$ is the magnitude of
the third brightest galaxy in the cluster.  Using our derived mass
luminosity relation, it is possible to transform this magnitude
threshold into a mass one.  In our model, this would correspond to the
mass of the second most massive subhalo (since we consider the
brightest galaxy to be in the parent halo).  In order to calculate
this, we assume that its mass is given by a distribution which is the
product of the Poisson probability that on average there exists a
single subhalo more massive than it, by the probability of finding a
subhalo at that mass. That is, its average value for a parent halo of
mass $M_h$ is given by

\begin{equation} \label{2ndmassavg}
<M_{sh,2}(M_h)>=\int_0^\infty M P_2(M,M_h) dM \, ,
\end{equation}

\noindent where $P_2(M,M_h)$ is the probability of the second most massive subhalo having mass $M$,

\begin{equation} \label{2ndmass}
P_2(M,M_h)=N(M|M_h) <N(M|M_h)> {\rm exp}(-<N(M|M_h)>) \, ,
\end{equation}

\noindent where $N(M|M_h)$ is the mass distribution function of the
subhaloes, equation \ref{nsubhalo}, and $<N(M|M_h)>=\int_M^\infty
N(M'|M_h) dM'$ is the average number of subhaloes more massive than
$M$ in a parent halo of mass $M_h$.

Since this mass threshold depends on the mass of the parent halo, in
the end we obtain a condition on halo mass for it to be treated as a
host to a rich cluster. Using our results, we find this to be
$M_h>2.85\times 10^{14} h^{-1} {\rm M_\odot}$.

If we now combine the distribution of all haloes more massive than
this, together with their subhaloes, with the mass luminosity
relation, we can obtain the luminosity function of galaxies in
clusters.  In fact, it is possible to make a rough prediction for its
shape at the high end by comparing the luminosity of the brightest
galaxy in the cluster (given by the solid line in figure
\ref{masslightfig}, which represents the luminosity of the galaxy
associated with the parent halo itself) with the total cluster
luminosity.  The average luminosity of the brightest galaxy can be
estimated as $L_1=\int_{L'}^\infty L \phi_{cl}(L) dL$, where
$\phi_{cl}(L)$ is the cluster galaxy luminosity function, and $L'$ is
such that on average there is only one galaxy more luminous than $L'$,
i.e. $\int_{L'}^\infty \phi_{cl}(L) dL=1$ (see
\citealt{YangMoBosch}). With $\phi_{cl}(L)$ normalized to the total
cluster luminosity, $L_{cl}$, it is then possible to extract a
relation between $L_1$ and $L_{cl}$, which will depend on the shape of
the luminosity function.  Taking our results from figure
\ref{masslightfig} that $L_1\propto M^{0.28}$ and $L_{cl}\propto
M^{0.9}$, it is possible to work out that, at the high end,
$\phi_{cl}(L)$ is given by a power law with slope -4.21.

\begin{figure}
\includegraphics[width=84mm]{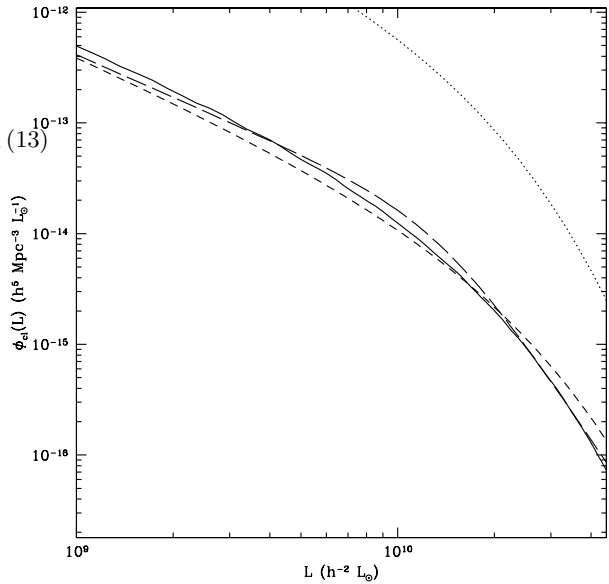}
\caption{Luminosity function of galaxies in rich clusters. Shown as
the solid line is our model, where we have excluded the first
brightest galaxies, and where we take an effective threshold of
$M_h>2.85\times 10^{14} h^{-1} {\rm M_\odot}$ for the mass of a halo
associated with a rich cluster. The dotted line shows the 2dF global
luminosity function, the short dashed one the luminosity function of
cluster galaxies also from 2dF, and the long dashed line is a double
power with the same parameters as the previous function, but with
$\phi_{cl}(L) dL \propto L^{-4.21}$ at high $L$.}
\label{clusterlffig}
\end{figure}

We show the results we obtain in figure \ref{clusterlffig}, where we
have excluded the first brightest galaxies (in effect, we are just
accounting for the galaxies in subhaloes). For comparison, we include
the 2dF global luminosity function of \citet{2dF}, as a dotted
line. We also include the luminosity function for galaxies in clusters
in the 2dF survey, as derived by \citet{2dfcluster}, as the short
dashed line. This is also given by a Schechter function, with
parameters $\alpha =-1.28$ and $M^*_{b_J}-5 {\rm log_{10}}
h=-20.07$. Since the normalisation is not given, we adjusted this to
have values in the same range as the ones we obtain. Also shown, as
the long dashed line, is a double power law with the same low end
slope, normalisation, and $L_*$ as the ones used to plot the previous
curve, but where the exponential cutoff has been replaced with a power
law with the same slope as calculated above, namely $-4.21$. Our
results show a good agreement to those obtained from 2dF, which is
even better if a double power law is considered instead of a Schechter
function.

\subsection{Group luminosity function}

Another way to check our results is to build the group luminosity
function.  This is the equivalent of the galaxy luminosity function,
only applied to groups, and it is usually obtained from galaxy
catalogues by building groups of gravitationally bound galaxies.  In
our case, we start with the same assumptions for groups described
above, and use the halo mass function and the relation between halo
mass and group luminosity given by equation (\ref{grouplum}) so that

\begin{equation} \label{grouplf}
\phi_g(L_g) dL_g= n_h(M(L_g))\frac{dM}{dL_g}dL_g \, .
\end{equation}

\begin{figure}
\includegraphics[width=84mm]{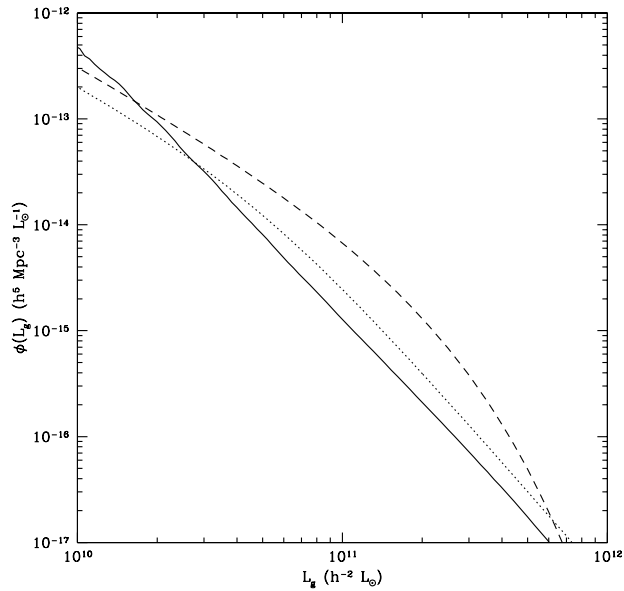}
\caption{Group luminosity function. The solid line represents our
result using equation (\ref{grouplf}); the two other lines are
observational group luminosity functions extracted from galaxy
catalogues, respectively the AGS \citep{AGS} (dotted) and the VSLF \citep{VSLF} (dashed).}
\label{grouplumfig}
\end{figure}

We show our result in figure \ref{grouplumfig}, along with the AGS for
the CfA survey \citep{AGS} and the VSLF \citep{VSLF} group luminosity
functions.  While at higher luminosities our model would seem to
underpredict the abundance of groups, it reproduces the slope of the
AGS luminosity function quite well.  This is undoubtedly related to
the fact that our calculated group luminosity is somewhat lower than
the observed values (see figure \ref{masslightfig}); an increase in
group luminosity would cause a shift to the right of our curve,
providing a better match for the observational results.  It should
also be noted that the inclusion of subsubhaloes would probably tend
to slightly increase the values at the high end, relatively to the low
end, since most of these subsubhaloes would be dark for low mass
parent haloes, but be more massive and therefore contribute a small
luminosity to the group total in the case of massive parent haloes.
It is also curious to note that, since the fits shown in the figure
are a double power law in the case of the AGS and a Schechter function
in the case of the VSLF, our result would probably not be well fit by
a Schechter type function.

\subsection{Multiplicity function}

\begin{figure}
\includegraphics[width=84mm]{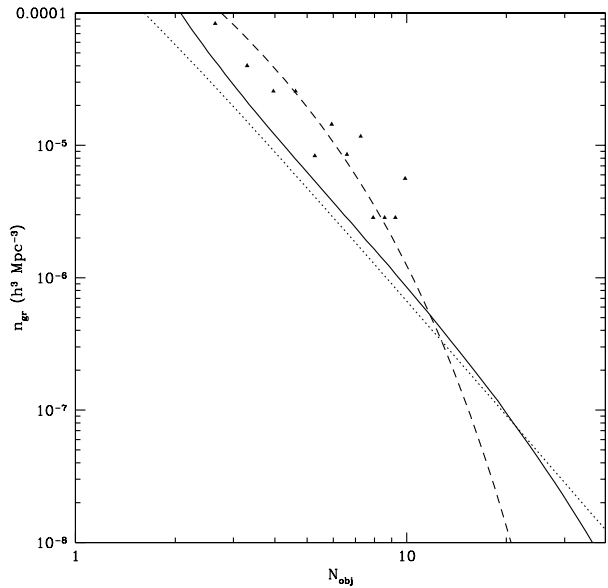}
\caption{Multiplicity function derived from our model (solid
line). This equates the number of galaxies in a group with the number
of subhaloes in a halo, plus one to account for a central galaxy
hosted by the halo itself. The minimum mass taken for the subhaloes
was in this case the equivalent to a magnitude of $M_B=-19.4$. The
different points are taken from the results derived by
\citet{PeacockSmith} from the CfA survey, while the two additional
lines are derived from the group luminosity functions shown in figure
\ref{grouplumfig}, where the magnitude limit for all of these is also
$M_B=-19.4$.}
\label{multfig}
\end{figure}

It is also possible to derive the multiplicity function,the number of
groups/clusters as a function of their richness, by a process similar
to the one used for the group luminosity function. Only in this case,
we use the occupation number (plus one to account for the central
galaxy, which is hosted by the parent halo itself) instead of the
group luminosity. Then, using an expression similar to (\ref{grouplf})
with the subhalo number $N_s$ put in place of the group luminosity
$L_g$, we obtain the result shown in figure \ref{multfig}. The main
point to take into account in this calculation is, as was referred in
our above discussion of the occupation number, the need to introduce a
minimum mass for the subhaloes. In the case of the figure shown, this
was taken to be the mass equivalent to $M_B=-19.4$, but in general,
and to compare to results from observational studies, this should be
set to equal the minimum luminosity considered for objects in the
observations.  There is also a sharp upturn at $N=1$ due to all the
haloes that are massive enough to be considered to host a galaxy, but
not enough so that they can have subhaloes with galaxies in them.

Figure \ref{multfig} shows our results in comparison with some
observational data. The points were taken from the analysis of
\citet{PeacockSmith}, while the two lines were constructed from the
two group luminosity functions shown in figure \ref{grouplumfig}, by
using the luminosity functions of the two surveys to relate total
group luminosity to the number of galaxies above the magnitude
threshold. The galaxy numbers for both the points and the results
derived from the group luminosity functions were then multiplied by a
factor of 0.66 to take into account the difference in radius between
friends of friends estimates and the usual definition of virial radius
(see \citealt{Kochaneketal}).

Even though our results lie in the range between those estimated from
the VSLF and the AGS group luminosity functions, they seem to be a bit
lower than the observational points. Since the group abundance is
directly related to the halo mass function, which is well known, this
discrepancy must be caused by the occupation number.  A slightly
higher occupation number would shift our curve further to the right in
the figure, bringing it into better agreement with the observational
data. Incidently, a higher occupation number would also bring our
results for the group luminosity and the related group lumininosity
function into better agreement with observational values, so we
believe that this is where the problem lies. Our calculated occupation
numbers seem to be in fair agreement with expectations; however, they
depend sensitively on our prescription for subhalo stripping, where we
adopted a simplistic approach. A more detailed and correct model for
this effect would most likely produce better results.

\section{Bias and probability functions}

\subsection{Mass, light and number densities}

To derive a relation between the mass and light densities, we first
need to obtain the mass function of haloes for regions of different
density.  In order to do this, we follow the method outlined in
\citet{Gottloeberetal}.  In essence, we treat the evolution of the
matter distribution in a region as if it derived from a universe with
the same cosmological parameters as that region, namely the same
average mass density as is found in the local region.  We start by
labelling each region by its average mass density, $\rho$, to which
corresponds a given value of the parameter $\Omega_m=\rho/\rho_c$,
with $\rho_c$ the critical density.  Overdense regions will have
$\Omega_m>\bar{\Omega}_m$, where the barred quantities refer to the
background universe, while voids will have the opposite.

The growth of perturbations will be affected by the different matter
content, and this is reflected by changing the normalisation of the
power spectrum according to the differential of the growth factor,
$D(z)$, relative to the background:

\begin{equation} \label{omegavoid}
\sigma_8=\bar{\sigma}_8\frac{\bar{D}(z_i)}{\bar{D}(0)}\frac{D(0)}{D(z_i)} \, ,
\end{equation}

\noindent where $z_i$ is some initial redshift for which the
fluctuations were equal in the background and in this region; in the
present case, we use $z_i=1000$. We also use the usual normalisation
for the growth factor, $\bar{D}(0)=1$.  To obtain the mass function
for a region with a particular average density, $n_{\rho}(m)dm$, we
then apply this new normalisation to the Sheth Tormen mass function
given in equation (\ref{stmf}), changing the $\rho_m$ term as
appropriate, and also the value for $\delta_c$, which has a small
dependence on $\Omega_m$ (see for example \citealt{NFW}).

\begin{figure}
\includegraphics[width=84mm]{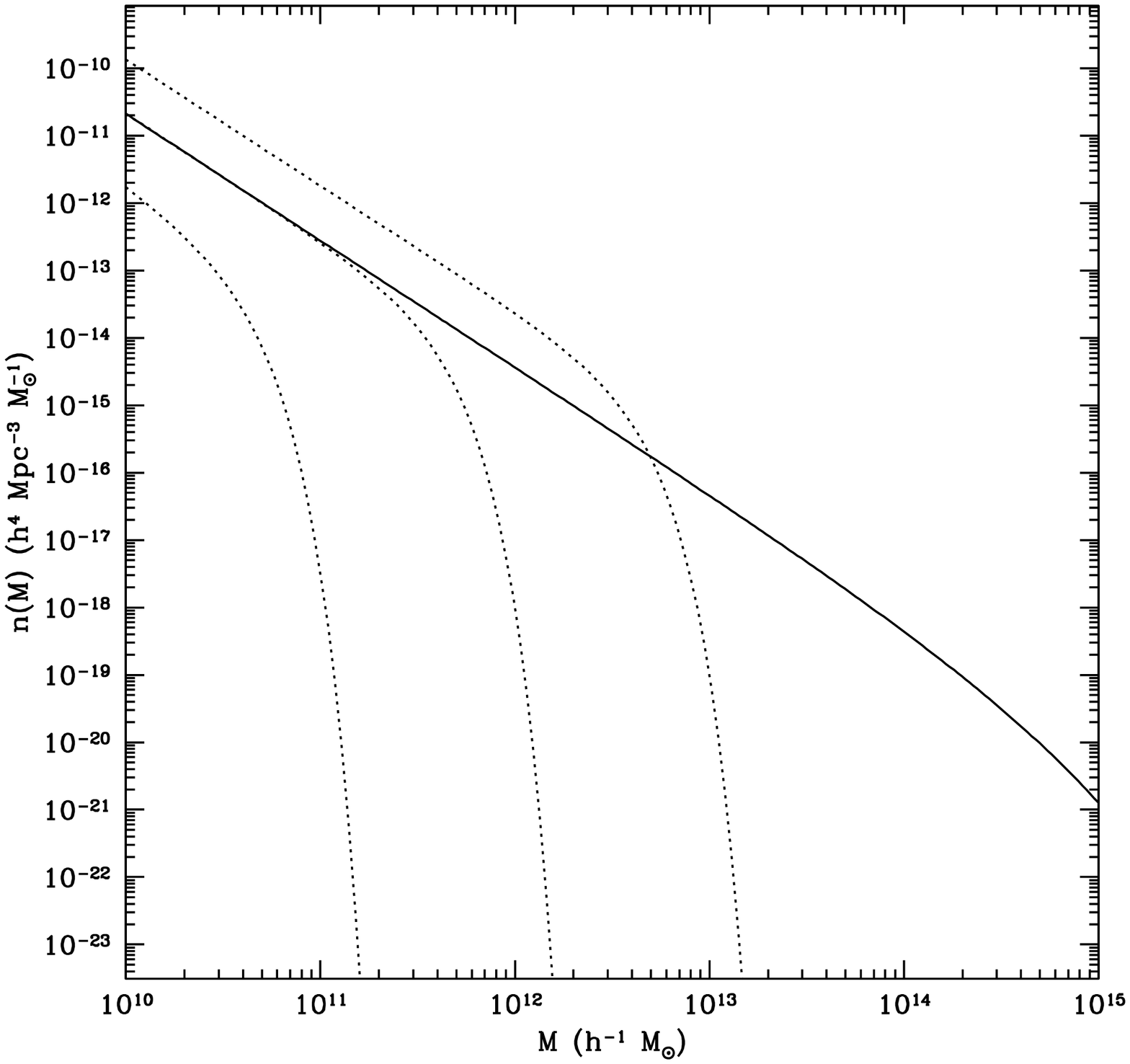}
\includegraphics[width=84mm]{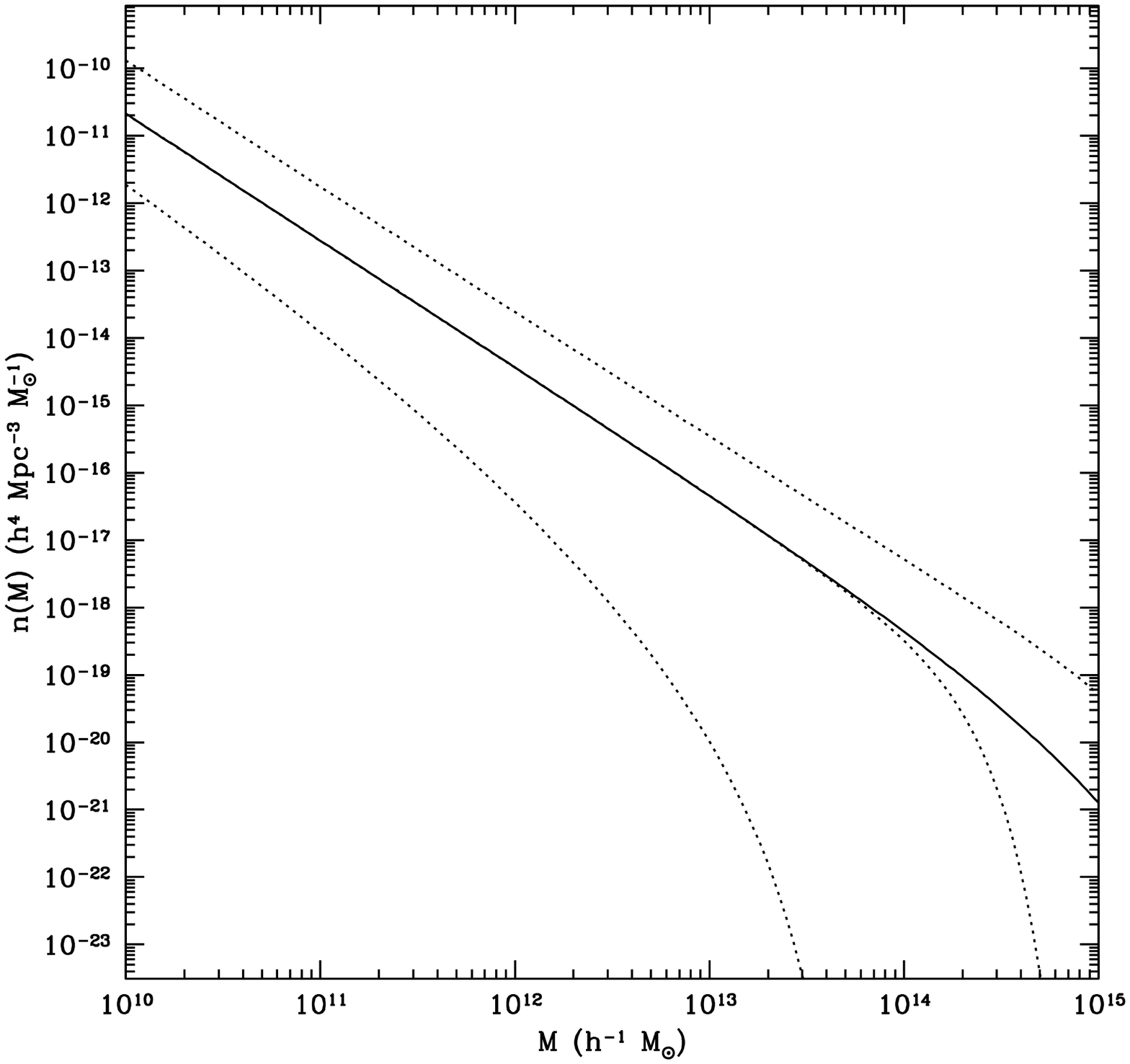}
\caption{Mass functions for regions of different size and average
density, calculated according to the prescription described in the
text. The upper panel shows the results for $R=1 h^{-1} {\rm Mpc}$,
the lower for $8 h^{-1} {\rm Mpc}$. In both, the solid curves are the
background Sheth Tormen mass function, the dotted ones are for
densities of(top to bottom): $10 \bar{\rho}$, $1 \bar{\rho}$, $0.1
\bar{\rho}$. }
\label{voidmffig}
\end{figure}

An important point is that this prescription does not utilise an
explicit smoothing radius for the region considered. Instead, these
are labelled by their average density.  However, when regions of
limited size are considered, there exists a maximum mass for objects
in them, given by the total mass they contain. As the above
prescription does not take this into account, we introduce an
additional term to compensate for this effect.  Therefore, after we
obtain the mass function, we put in a further  cutoff of the form
${\rm exp}[-\eta (m/m_\delta)^2]$, where $m$ is the mass of the halo
and $m_\delta=4\pi (1+\delta_m) \bar{\rho} R^3 /3$ the total mass in
the region of radius $R$ and average density $(1+\delta_m)
\bar{\rho}$. The parameter $\eta$ allows for some tuning of the actual
cutoff, and in the present case we use $\eta=1$.  Once this cutoff is
introduced, it becomes necessary to renormalise the mass function, so
that it still gives the appropriate density when the mass of all
haloes is calculated. Two examples of the resultant mass functions are
shown in figure \ref{voidmffig}, with curves for regions with the
average density, $1/10$ of the average and $10$ times the average, for
two different radii.

Once the mass function for an over or under dense region has been
determined, it is then simple to find the equivalent light density,
$\rho_L$, by using the group luminosity correspondent to each halo,
given by equation (\ref{grouplum}):

\begin{equation} \label{rholight}
\rho_L=\int_0^{\infty}L_g(M) n_h(M) dM \, .
\end{equation}

\begin{figure}
\includegraphics[width=84mm]{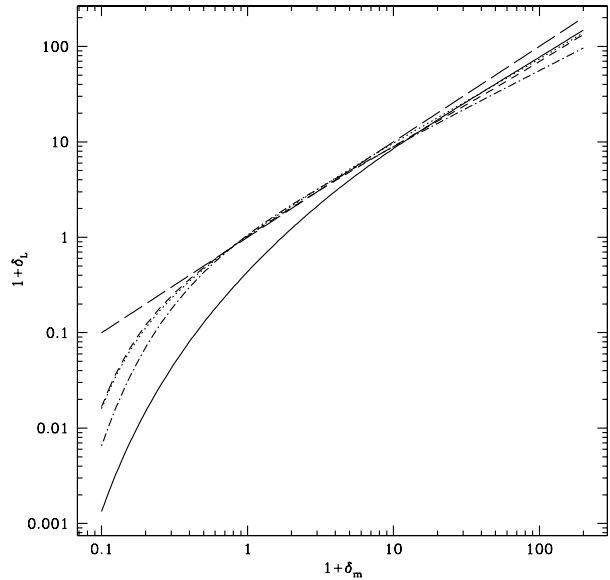}
\caption{Light density, $\delta_L$, as a function of mass density,
$\delta_m$. Both axis are scaled to the average background values, so
regions with $1+\delta_m>1$ are overdense and regions with
$1+\delta_m<1$ are underdense. The different lines are for different
smoothing lengths: $R=1 h^{-1} {\rm Mpc}$ (solid), $R=4 h^{-1} {\rm
Mpc}$ (dotted), $R=8 h^{-1} {\rm Mpc}$ (short dashed). The long dashed
line marks $\delta_L=\delta_m$. The dashed dotted line is the fit to
the same relation mentioned in \citet{Moetal}.} \label{mldensityfig}
\end{figure}

Our result is shown in figure \ref{mldensityfig}. The results for $R=8
h^{-1} {\rm Mpc}$ and $R=4 h^{-1} {\rm Mpc}$ are quite similar, with
the curve for $R=1 h^{-1} {\rm Mpc}$ being different. This is due to
the large suppression of high mass haloes, even at high overdensities,
which can be seen in figure \ref{voidmffig}, which alters the
proportion of luminous to non-luminous haloes in favour of the latter.
The curve we obtain for $R=8 h^{-1} {\rm Mpc}$ is similar to the one
in \citet{Moetal}, where the authors have used a mass function for
different densities derived from simulation results, to which they
then apply the conditional luminosity function. The agreement is quite
good with our result, even though we used a theoretical model for the
mass function instead of taking it from simulations.

The most striking feature in these results is the sharp decline in the
light density in underdense regions. This would mean that the majority
of void regions would be quite dark.  Another important feature is
that for the most part $\delta_L<\delta_m$, and the slope of the curve
at high density is lower than 1.  The overall results (e.g., darkness
of the voids) are similar to those from  hydrodynamical simulations by
\citet{OstrikerNagamineFukugita}.  But the simulation results show a
sharper cutoff at low mass density and light being more overdense than
mass for overdense regions and a slope to the relation greater than
one.  This difference is the more remarkable since the results of
\citet{Moetal} are similar to ours while using a mass function derived
from simulations, which otherwise might be considered the most likely
origin of the discrepancy.  However, previous results shown in
\citet{bcdoy}, also taken from simulations, seem to be more in line
with what we obtain in the present work.  These authors in fact give
an explanation for the apparent increasing antibias at higher mass
densities: at low redshift these regions usually consist of rich
clusters and superclusters, whose stellar population tends to be
old. Therefore, young blue stars tend to be rare and consequently the
total luminosity in the blue band is lower than what could be expect
from their high mass, giving rise to the slight antibias.  It is
therefore likely that this difference is coming from the way in which
light is being counted in the two methods.  In fact, one problem with
the approach presented here is that it breaks down for haloes whose
size is comparable to the size of the region being considered, since
it is then possible for a sphere to encompass an outer region of the
halo where a subhalo and a galaxy lie, and which is therefore luminous
but is considered dark in our prescription.  Since
\citet{OstrikerNagamineFukugita} study the light distribution
directly, they would account for the light present in such a
situation.  This effect would be the more noticeable for high density
regions and small smoothing lengths, and in fact the results presented
in \citet{OstrikerNagamineFukugita} are for $R=1 h^{-1} {\rm Mpc}$,
while the major discrepancy with our results occurs for high density
regions. The differences in underdense regions can most likely be
explained by limited resolution effects.

\begin{figure}
\includegraphics[width=84mm]{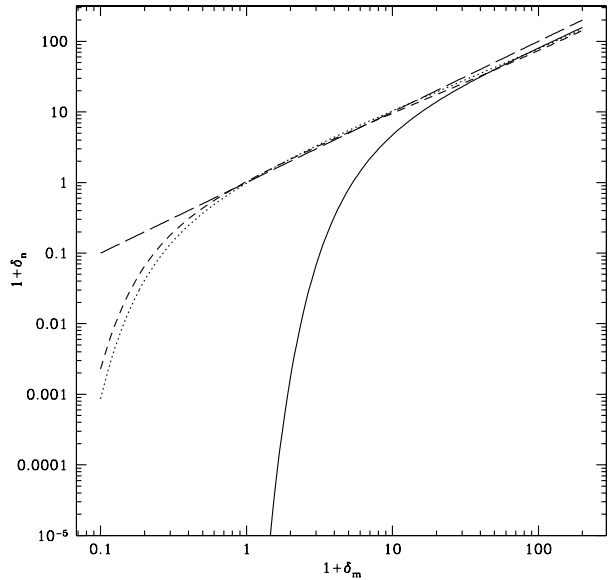}
\caption{Same as figure \ref{mldensityfig}(curves have same
label), but for $\delta_g$, the overdensity in the number of
galaxies. The minimum threshold was set at halo mass corresponding
to a luminosity $L=L_*$. The average number of galaxies brighter
than this limit in the background, taken from the luminosity
function, is $3.22\times10^{-3} h^3 {\rm Mpc}^{-3}$.}
\label{mndensityfig}
\end{figure}

We also derive results for the bias between galaxy numbers and the
dark matter mass density. The procedure is similar to the adopted for
the luminosity, replacing the $L_g(M)$ term in (\ref{rholight}) with
the number of galaxies in a halo of mass $M$,
i.e. $(1+N_s(M,m_{min}))$, corresponding to the central galaxy hosted
in the halo and the ones present in the subhaloes. The integral should
also be taken with a lower limit at $m_{min}$. The minimum mass
$m_{min}$ corresponds to the intended cutoff in mass (or luminosity)
for the galaxies to be counted. Our result is shown in figure
\ref{mndensityfig}. The general behaviour is similar to what was
observed for the luminosity. According to these results, the galaxy
number distribution would be a fairly unbiased tracer of the
underlying dark matter mass in moderately overdense regions, as would
be expected from previous studies (e.g., \citealt{bias}), but the
galaxy numbers severely underestimate the mass density in underdense
regions and consequently assuming a direct proportionality can lead to
a significant underestimate of $\Omega_m$
\citep{OstrikerNagamineFukugita}.

\subsection{Distribution functions}

Using the relations obtained in the previous section, it is also
possible to derive a distribution function for the light density
(i.e., the probability of a region of a given size having a certain
average light density, the light PDF). To make this calculation,
besides the relation between the mass and light densities it is also
necessary to have the mass distribution function.  In general, this
can be obtained from dark matter simulations. For the present work, we
will use the approximation of treating this as a lognormal, with an
appropriate dispersion dependent on the radius being considered and on
the cosmological model. It has been known for some time that such a
function is a fair approximation to the real mass distribution
function, at least in regions of moderate over/under density
(\citealt{ColesJones}; see also \citealt{Kayoetal} and
\citealt{OstrikerNagamineFukugita} for some recent analysis on this
subject).

The form of the lognormal is given by:

\begin{equation} \label{lognormal}
f(y)=\frac{1}{\sqrt{2\pi \omega^2}}\frac{1}{y}{\rm exp}
\Big(-\frac{[{\rm ln}(y)+\omega^2/2]^2}{2\omega^2}\Big) \, ,
\end{equation}

\noindent where $y=1+\delta_m$ and the parameter of the distribution
$\omega$ is related to the smoothing length and the variance $\sigma$
of the mass density field by

\begin{equation} \label{lnpar}
\omega_{\rm R}^2={\rm ln}(1+\sigma_{\rm R}^2) \, .
\end{equation}

\noindent Here, the variance $\sigma$ refers to the full, nonlinear
spectrum. To relate this to the linear variance, and hence to the
cosmological parameters we use, we follow the prescription outlined in
\citet{PeacockDodds} to convert the linear power spectrum of mass
density fluctuations to the nonlinear one.

\begin{figure}
\includegraphics[width=84mm]{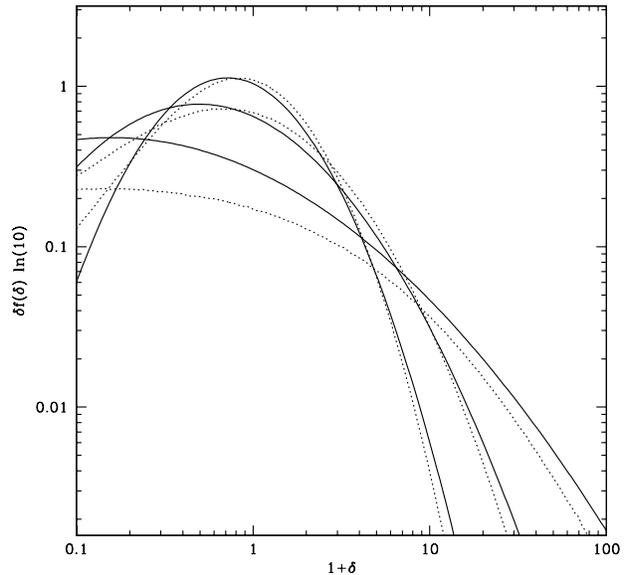}
\caption{Distribution functions for mass (solid) and light (dotted)
density. The three curves correspond to the three smoothing lengths
used previously: 1, 4 and 8 $h^3 {\rm Mpc}^{-3}$, with peak height
increasing with radius. The x-axis represents either mass or light
overdensity, depending on the curve. The distribution functions are
shown scaled appropriately for logarithmic binning.}
\label{distfuncfig}
\end{figure}

To obtain the light distribution function, we then follow a similar
procedure to the one used for the group luminosity function:

\begin{equation} \label{lightpdf}
g(j) dj= f(\rho)\frac{d\rho}{dj}dj \, ,
\end{equation}

\noindent where we use the notation $j=1+\delta_L$ and $g(j)dj$ is the
light distribution function. Such a transformation assumes a monotonic
dependence of the light density with the mass density, which is true
of our results (see figure \ref{mldensityfig}). Another important
point to note is that this function is not normalised to one, since
some of the underdense areas (in terms of mass) are non luminous (see
discussion in \citealt{OstrikerNagamineFukugita}). Our result is shown
in figure \ref{distfuncfig}. The general behaviour is what would be
expected: the distribution functions are similar in the overdense
regions, but in general differ substantially in the underdense
regions, with the light distribution function in general not well fit
by a lognormal.

\subsection{Void probability function}

The void probability function is defined as the probability of having
no galaxies in a sphere of radius $R$. Although voids are a
particularly striking feature in the galaxy distribution, their study
is impaired by their large size and low number of galaxies present in
the observational case, and by resolution difficulties when the study
is conducted through numerical simulations including galaxy
formation. Due to these factors, in the past studies of the voids in
the galaxy distribution have been relatively few (see, for example,
\citealt{voidsobs} for observational results, and \citealt{voidmathis}
and \citealt{voidbenson} for comparisons with theoretical results;
there are also two recent papers with observational results for the
2dF survey, \citealt{hv,void2df}). Despite this, the void probability
is a powerful tool in the analysis of a particular model of galaxy
formation, since it probes a highly non-linear regime and is not
derivable from the low order correlation functions (in fact, it
depends on all of the N-point correlation functions).

In the case of our model, we can determine the void probability
function by combining the dark matter distribution function with the
expected galaxy number in regions of a given size. Thus the number
density of galaxies in a region of radius $R$ and average density
$\rho$, given a minimum mass cutoff of $M_{min}$ will be:

\begin{equation} \label{avgno}
n_R(\rho)=\int_{M_{min}}^\infty (1+N_s(M))n_{\rho,R}(M)dM \, ,
\end{equation}

\noindent where $n_{\rho,R}(M)$ is the mass function in a region of
size $R$ and average density $\rho$ (see discussion in section 4.1,
also figure \ref{voidmffig}), and $N_s(M)$ is the number of subhaloes
in a halo of mass $M$. The number is then obtained by multiplying this
density by the region volume, $V=4\pi R^3/3$. We treat this as the
average number and assume a Poissonian distribution around this
average (see \citet{Kravtsovetal}, where the authors have shown that
the full HOD is consistent with a Poisson distribution for large host
masses), so that the probability of having 0 galaxies will in fact be
${\rm exp}(-V n_R(\rho))$. Finally, the void probability function is
obtained by integrating this number times the probability of a region
having this given density,

\begin{equation} \label{voidpf}
P_0(R)=\int_0^\infty {\rm exp}(-V n_R(\rho)) f_R(\rho) d\rho \, ,
\end{equation}

\noindent where $f_R(\rho)$ is the probability of a region of radius R
having density $\rho$, which we take to be given by the lognormal
distribution, equation (\ref{lognormal}).

\begin{figure}
\includegraphics[width=84mm]{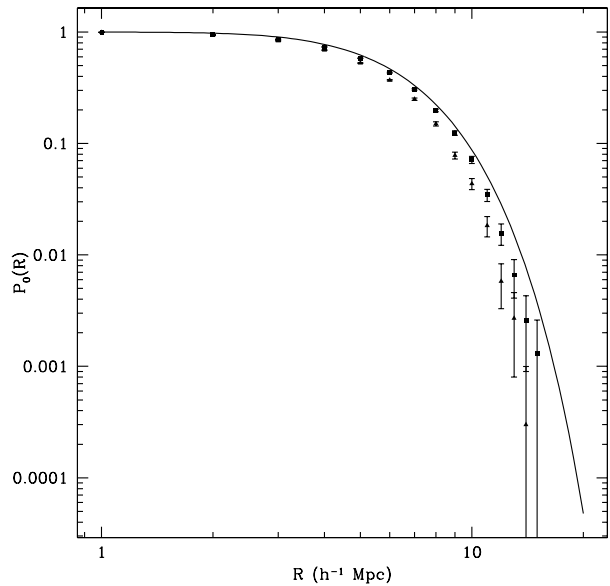}
\caption{Void probability function. The solid line represents our
concordance model.  The points are from the combined CfA-1 and CfA-2
surveys (taken from \citealt{voidbenson}, original results in
\citealt{voidsobs}). The observations are for a magnitude limit of
$M_B-5{\rm log}h\le-19.5$, and the corresponding mass was taken as the
minimum mass limit in the calculations (equation \ref{avgno}).}
\label{voidpffig}
\end{figure}

Our results are shown in figure \ref{voidpffig}, where we also show
observational results for the 2dF survey taken from \citet{hv}, for
both the NGP and the SGP. At low radius, we obtain a good agreement,
but our model would seem to overestimate the probability at large
radii. Our results are also in better agreement with the values
observed for the SGP than for the NGP, since \citet{hv} have found the
former to be somewhat emptier than the latter.  There is an additional
effect that may go some way towards explaining this discrepancy, and
that is corrections for the peculiar velocity distortions. These would
lead, especially for the larger radii, to the inclusion in voids of
galaxies that are not actually there, due to the smearing effect
caused by peculiar velocities to the positions of galaxies in redshift
space. This effect is not symmetric, since the voids are areas of low
mass density and therefore the galaxies in them will have low velocity
dispersions. The final result of this would be to overpopulate the
voids and consequently to underestimate the measured void probability
function, which would lead to a better agreement with our results.

\section{Conclusion}

In this paper we present a new model for relating halo mass to hosted
galaxy luminosity, based on the dark matter substructure.  We feel
this is a potentially powerful way to approach this problem, since it
is based on two main inputs, the halo/subhalo distribution and the
galaxy luminosity function, which can be well tested and adjusted to
results from simulations and surveys, respectively.  Additionally, the
model requires only one further assumption: that there is a one to one
and monotonic relation between imbedded halo/subhalo mass and galaxy
luminosity. This is more explicit but in general agreement with the
general assumptions made in past studies and general assumptions made
in similar work of the same subject. It is a much less restrictive
assumption than the still sometimes utilized ansatz of a linear
``bias'' between galaxy numbers and dark matter density.

We have shown how, starting with a prescription we describe for the
subhalo mass distribution in a parent halo, it is possible to obtain a
relation for the luminosity of a hosted galaxy, as well as the group
luminosity when the system of a halo and its subhaloes is identified
with groups or clusters of galaxies. The subsequent results appear to
match well with general assumptions of the behaviour of such a
relation, as well as results for the mass and luminosity at scales of
dwarf galaxies, $L^*$ galaxies and massive clusters.

From this model, it is then possible to derive many quantities that
can be directly compared with further simulation or observational
results. We have shown four examples of this, namely the occupation
number, the luminosity function of cluster galaxies, the group
luminosity function and the multiplicity function.  Using further
assumptions on how to calculate the mass function for regions of
different average densities, and on the shape of the dark matter
distribution function, we have also obtained a relation between mass
and light and number densities for different smoothing lengths close
to what is expected from previous bias studies. We also obtain the
distribution function of light density, and the void probability
function. The latter is a powerful additional test, since it probes a
highly non-linear regime, and our results seem to match well with
previous observations.

The major difficulty with the model as it is presented here has to
do with the identification of the mass of the subhaloes. To be able
to apply the monotonic correspondence to the galaxy luminosity,
the original mass in the subhaloes has to be taken into account.
However, the mass distribution we use is measured for the stripped
mass of the subhaloes in their parent halo. In order to build the
model, we took the approximation of taking an average for the
stripping factor. This is, however, a not wholly satisfactory
approach, since in general the stripping history will be highly
variable from subhalo to subhalo, and also in different parent
haloes. Ideally, we would like to use the maximum circular velocity
instead of the mass to identify the subhaloes, since this quantity
should less sensitive to stripping for the relatively massive
subhaloes which can host galaxies. Unfortunately, we do not have at present 
a good distribution for subhalo abundance as a function of the
maximum circular velocity, although we wish to study this further in 
the future. Nonetheless, although somewhat crude,
the approximation we took is reasonable, and since this work deals
with statistical averages for most of the quantities, it does fit
somewhat well into the whole structure. More importantly, the
results we obtain seem for the most part to match well with
observational data. In future work which takes into account the
statistical variation, however, this factor will certainly be of
great importance.

\section*{Acknowledgements}
We wish to thank Michael Strauss and Jochen Weller for useful
discussions. We are also very grateful to the anonymous referee for a
most helpful report. AV acknowledges financial support from Funda\c
c\~ao para a Ci\^encia e Tecnologia (Portugal), under grant
SFRH/BD/2989/2000.

\end{document}